\documentclass[%
 reprint,
 amsmath,amssymb,
 aps,
]{revtex4-2}

\usepackage{graphicx}
\usepackage{dcolumn}
\usepackage{bm}
\usepackage[caption=false]{subfig}
\usepackage{xcolor}

\begin{document}

\preprint{APS/123-QED}

\title{First-order and Berezinskii-Kosterlitz-Thouless phase transitions in two-dimensional generalized XY models }% Force line breaks with \\

\author{P. A. da Silva}
\author{A. R. Pereira}
\affiliation{Departamento de F\'{i}sica, Universidade Federal de Vi\c cosa, Avenida Peter Henry Rolfs s/n, Vi\c cosa, MG, 36570-900, Brazil.}
\author{R. J. Campos-Lopes}
\affiliation{International School for Advanced Studies (SISSA), via Bonomea 265, 34136 Trieste, Italy}

\begin{abstract}
{\color{black} Besides the Berezinskii-Kosterlitz-Thouless phase transition, the two-dimensional generalized XY model, identified by a generalization parameter $q$ (as proposed by Romano and Zagrebnov), can also support a first-order phase transition, starting from a critical value $q_c$. However, the value of $q_c$ at which this transition takes place is unknown. In this paper, we take two approaches to determine the critical parameter $q_c$ accurately.} Furthermore, we show that the model is characterized by three distinct regions concerning both first-order and Berezinskii-Kosterlitz-Thouless phase transitions. Finally, the underlying mechanism governing such transitions is presented, along with an estimation of the critical temperatures.
\end{abstract}

\maketitle

\section{introduction}

\begin{figure*}
\begin{minipage}{.32\textwidth}
    \subfloat[]{\includegraphics[width=\textwidth]{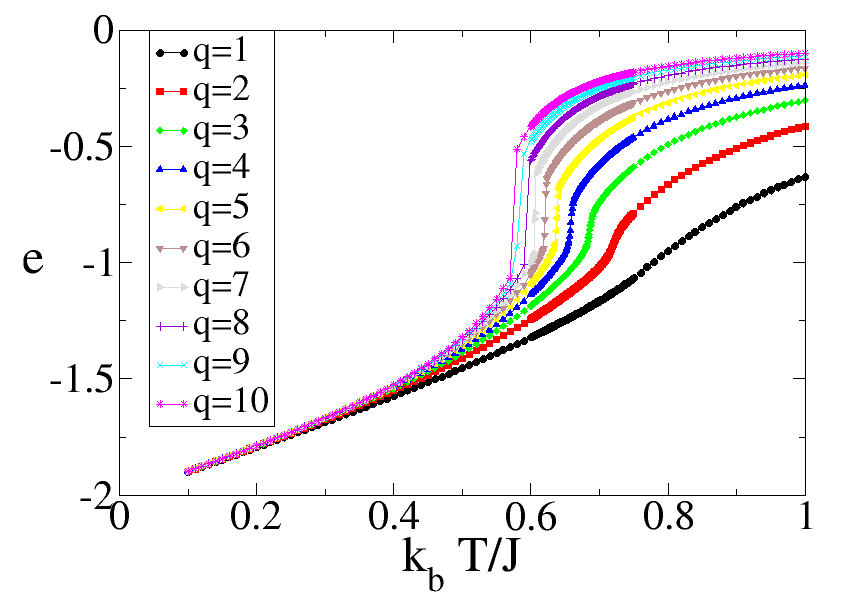}\label{Energia}}
\end{minipage}
\hfill    
\begin{minipage}{.32\textwidth}
    \subfloat[]{\includegraphics[width=\textwidth]{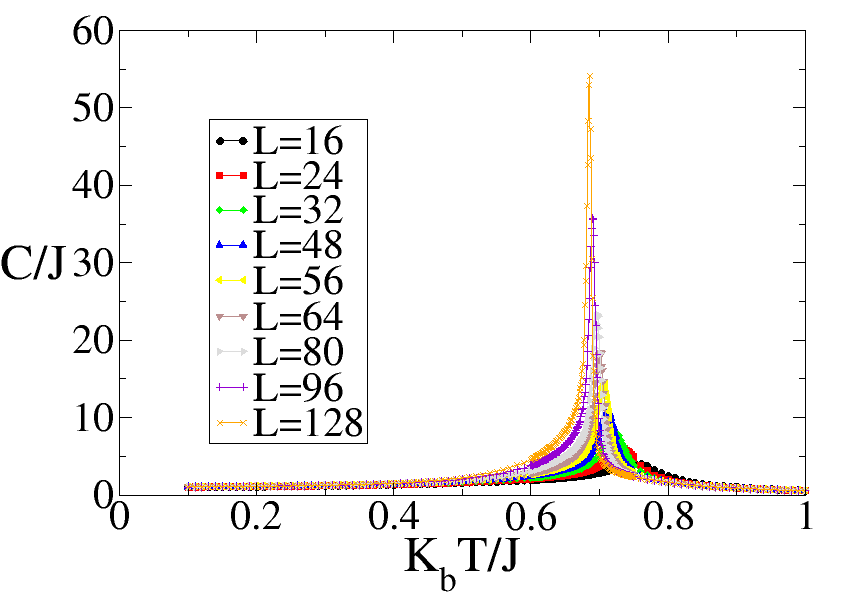}\label{fig:sub_a}}
\end{minipage}
\hfill    
\begin{minipage}{.32\textwidth}
    \subfloat[]{\includegraphics[width=\textwidth]{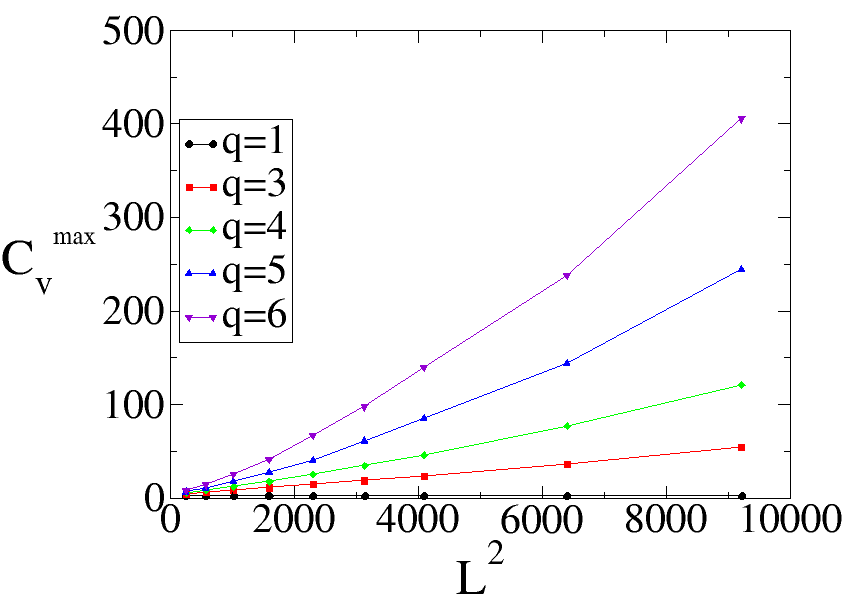}\label{Cvmax}}
\end{minipage}
    \caption{(a) Energy per spin as a function of temperature for different values of $q$ and $L=96$. As $q$ increases, we can observe a discontinuity being formed. Specific heat for $q=3$ (b), and the scaling of max specific heat with $L^{2}$ (c), showing the {\color{black} behavior} of a first-order phase transition.}
\end{figure*}

Two-dimensional magnetic models characterized by short-range interaction and continuous symmetry fail to exhibit {\color{black} magnetic long-range order} at a finite temperature \cite{mermin1966absence}. Despite the absence of {\color{black} magnetic long-range order} due to the destructive influence of spin waves, even at low temperatures, a quasi-long-range order with a power-law decaying correlation function {\color{black} may occur}. At higher temperatures, correlations {\color{black} always decay} exponentially, leading to a disordered phase. The transition from quasi-long-range order to a disordered phase constitutes a topological phase transition within the Berezinskii–Kosterlitz–Thouless (BKT) universality class \cite{berezinskii1971destruction,berezinskii1972destruction}. This transition is characterized by the unbinding of topological objects, known as vortices, at a critical temperature $T_{BKT}$.

Among the simplest models supporting topological excitations in the spin field and a BKT transition are those described by the Hamiltonian $H= -J \sum_{<ij>} (S^{x}_{i} S^{x}_{j} + S^{y}_{i} S^{y}_{j})$, where $J$ is a coupling constant, the summation is over nearest-neighbors sites in a square lattice, and $S^{\alpha}_{i}$ represents the spin at site $i$. From this Hamiltonian one can extract the Planar Rotator Model (PRM) by enforcing the constraint $S^{2}_{x} + S^{2}_{y} = 1$, which confines the spins to the plane. On the other hand, this same Hamiltonian leads to the XY model, simply by allowing spins to have three components and the constraint $S^{2}_{x} + S^{2}_{y} + S^{2}_{z} = 1$.  Both models can be parameterized by using scalar fields. For the PRM, the azimuthal angle $\phi$ is enough, yielding the Hamiltonian $H=-J \sum_{<ij>} \cos( \phi_i - \phi_j)$. Meanwhile, for the XY model, both azimuthal and polar angles are employed, resulting in the Hamiltonian $H=-J \sum_{<ij>} \sin \theta_i \sin \theta_j \cos( \phi_i - \phi_j)$. 

Extensive studies on the static critical behavior of such models have established consensus regarding phase transitions, critical temperature, and exponents \cite{nho1999critical} \cite{krech1999spin}. The nature of the phase transitions was {\color{black} intensively} investigated in the 1970s and 1980s, such that different generalizations and extensions of these two models have become objects of study.

Domany et al. \cite{domany1984first} introduced a generalization of the PRM model utilizing the Villain model $V(\phi)= 1 - \cos^{p^2}(\phi'/2)$, with $\phi'=\phi_{i} - \phi_{j}$.
As p increases, a narrow potential well with width $\pi / p$ emerges, resembling the n-states Potts model with p proportional to n. With Monte Carlo simulations, Domany et al. \cite{domany1984first} revealed that an increase in p leads to a discontinuity in energy and a two-peak structure in the energy histogram, suggesting a first-order transition for large p. Some studies challenged this, particularly those based on renormalization group analysis \cite{knops1984first,jonsson1993new}. However, the possibility of first-order transition {\color{black} between disordered phases} induced in the planar model was demonstrated \cite{korshunov1992disorder}, and later rigorous proof by Van Enter and Shlosman confirmed that SO(n) invariant $n$-vector models with sufficiently deep and narrow minima can undergo a first-order phase transition\cite{van2002first}. %{\color{black} Further, concerning the nature of the involved phases in a first-order phase transition (for models with continuous symmetries), a slightly more extensive discussion can be found in Ref.\cite{van2005second}.}

A generalized XY model was proposed by Romano and Zagrebnov\cite{romano2002xy}, whose Hamiltonian is given by,

\begin{eqnarray}
H^{Gen}_{XY} =&&-J \sum_{<i,j>} \big[ 1 - (S^{z}_{i})^{2} - (S^{z}_{j})^{2} \nonumber\\
&&
+ (S^{z}_{i} S^{z}_{j})^{2} \big]^{(q-1)/2} \left( S^{x}_{i} S^{x}_{j} + S^{y}_{i}S^{y}_{j} \right).
\end{eqnarray}
Writing this with the help of the two scalar fields $\theta$ and $\phi$, one gets

 \begin{equation}
    H^{Gen}_{XY} = -J \sum_{<i,j>} ( \sin \theta_i \sin \theta_j )^{q} \cos(\phi_{i} - \phi_{j}) .
\end{equation}
Here, $q$ is the generalization parameter and the XY model is recovered for $q=1$. {\color{black} Although $q$ does not need to be an integer number for the Hamiltonian model to be well-defined, in this work, we will consider only integer numbers for $q$ as defined by the original articles}. In their paper\cite{romano2002xy} they used rigorous inequalities for all values of $q$ for systems in two and three dimensions. For the two-dimensional (2D) case, they showed that, for arbitrary $q$, the model has orientational disorder at all finite temperatures, and undergoes a transition to a low-temperature phase with slow decay of correlations and infinite susceptibility. {\color{black} In thermodynamic terms, }the class of universality of the transition {\color{black}may or may not correspond to that of the proper BKT transition.} M\'{o}l \textit{et al.} \cite{mol2003phase} extensively studied the vortex-like solutions within the generalized XY model, employing both the continuous Hamiltonian and the Self Consistent Harmonic Approximation (SCHA) \cite{pereira1995phase}. They showed that only static planar-vortex configurations are stable and supported for any $q \geq 1$ and that the critical temperature of phase transitions decreases with the parameter $q$. In addition, Monte Carlo simulations \cite{mol2006monte} suggest a first-order phase transition in the generalized XY model {\color{black}for sufficiently high values of $q$}, but the result was not conclusive. {\color{black} Although} the first-order transition was initially contested, a rigorous proof by van Enter et al. \cite{van2006first} confirmed that the 2D generalized XY model proposed by Romano and {\color{black} Zagrebnov} indeed exhibits a first-order phase transition, leaving open the value of the critical parameter $q_c$ at which this transition begins to appear. %{\color{black}  So, concerning the perspective of a first-order phase transition in this generalized XY model, further important steps are associated with the value of $q_c$.}

{\color{black} Above, we have seen part of the effort to understand the generalized XY model. However, there are still several open questions regarding the two different phase transitions and their possible coexistence in the system. Here, our main motivation is to establish the correct values of the generalized critical parameter $q_c$, above which a first-order phase transition takes place. It is a fundamental step for further investigations related to the model.} For that, we have used the hybrid Monte Carlo algorithm, not much different from that used by Mól et al. \cite{mol2006monte}, which will be explained briefly in the next section. To enhance readability, most thermodynamic results and techniques for determining transition temperatures are presented in appendices, with only the main results in the usual text. We begin by presenting Monte Carlo simulation results and then, guided by these calculations and the solution of van Enter et al. \cite{van2006first}, we analytically estimate the critical temperature of the first-order phase transition.

\section{Monte Carlo Method}

\begin{figure*}
\begin{minipage}{.32\textwidth}
    \subfloat[]{\includegraphics[width=\textwidth]{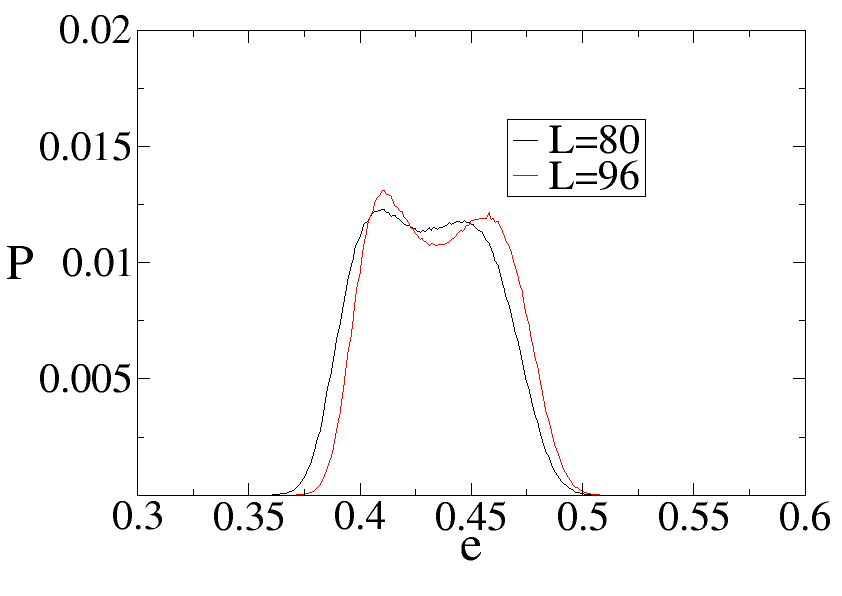}}\label{Histq3}
\end{minipage}
\hfill    
\begin{minipage}{.32\textwidth}
    \subfloat[]{\includegraphics[width=\textwidth]{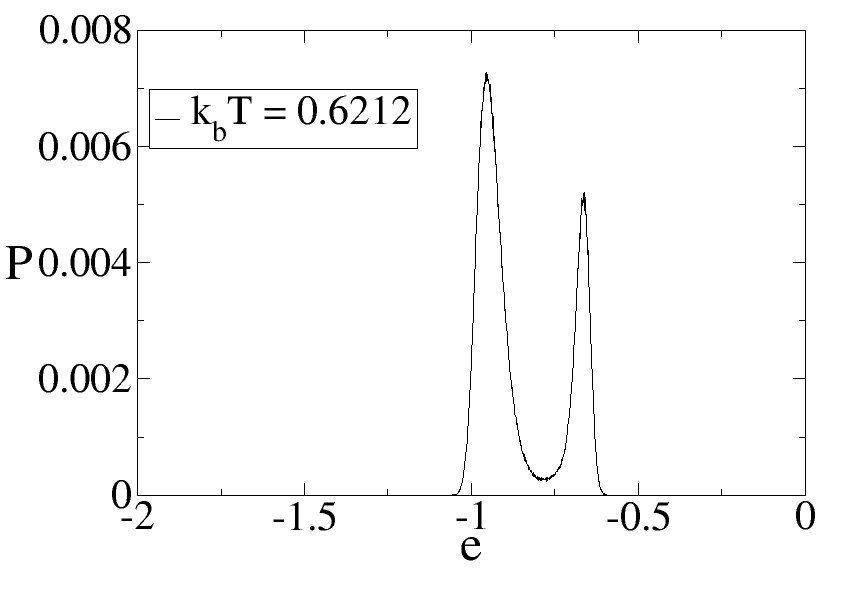}}\label{Histq6}
\end{minipage}
\hfill    
\begin{minipage}{.32\textwidth}
    \subfloat[]{\includegraphics[width=\textwidth]{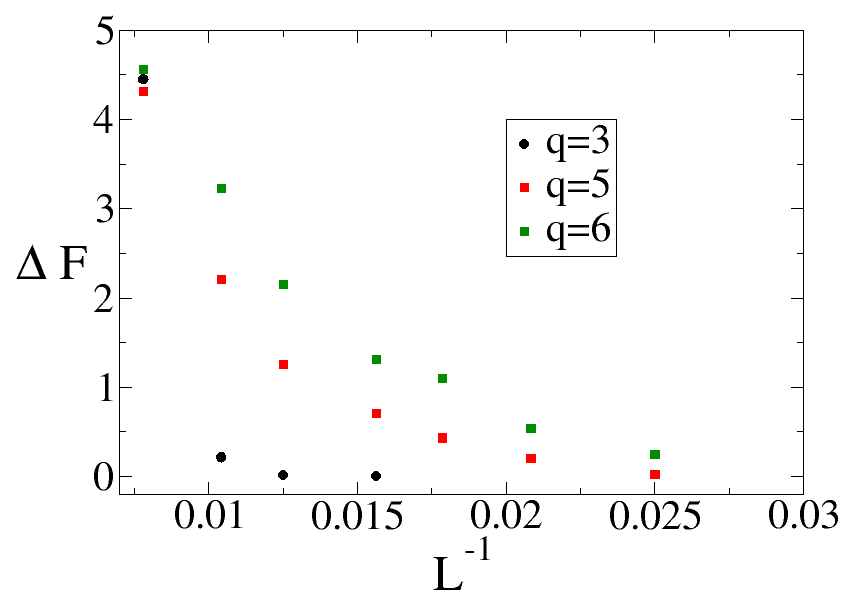}\label{LeeKosterlitz}}
\end{minipage}
\caption{(a) shows how the gap varies with the size of the system, comparing the two sizes of $L=80$,$96$.(b) the two peak-structure is easily observed for $q=6$ and $L=96$ at the transition temperature. (c) Lee-Kosterlitz criterion for $q=3$,$5$,$6$, where the expected {\color{black} behavior} of a first-order transition is observed for the three cases.} \label{q3Histogram}
\end{figure*}

We have employed a Hybrid Monte Carlo approach to generate configurations for calculating thermal averages. A Monte Carlo Step (MCS) involves a sequence of operations, starting with a Wolff cluster algorithm for in-plane components \cite{wolff1989collective}, followed by modified Metropolis single-spin updates for the three components and N over-relaxation steps. In the modified Metropolis algorithm \cite{robert2004metropolis}, a randomly selected spin underwent a small increment in random components, followed by renormalization to maintain unit length. The acceptance was determined by the standard Metropolis algorithm, with the spin increment length chosen to achieve an acceptance rate between 30\% and 60\%. This modification proved effective in mitigating critical slowdown at low temperatures. Several acceptance rates were tested, with little impact on the results. Over-relaxation involved reflecting a randomly selected spin across the effective magnetic field due to its neighbors while preserving the spin length. This algorithm effectively altered configurations while maintaining energy conservation. Mean values were obtained using 10,000 Monte Carlo passes to warm up the samples, and 300,000 configurations were individually divided into each bin for analysis. Testing higher numbers of configurations and bin sizes has shown minimal impact on the results.

\section{Results}

\subsection{Monte Carlo Results}

Initially, we will examine the impact of the generalization parameter $q$ on the energy for spin ($e$) and the specific heat ($c_{v}$).  For the simulations involving the usual XY model ($q=1$), our results align with expectations \cite{cuccoli1995two}. Both energy and specific heat exhibit minor finite-size effects as we approach the critical temperature; the estimated errors escalate due to rising fluctuations and critical slowing down. A distinct peak emerges at a temperature exceeding the critical temperature of the BKT transition. The determination of this critical temperature involves various methods, detailed in table \ref{TabCrit} and further explained in the appendix. With the increasing of the parameter $q$, the specific heat peak moves toward lower temperatures (still above the critical temperature observed for a BKT transition) and becomes narrow and higher as the system size $L$ increases. The energy begins to exhibit a discontinuity (fig. \ref{Energia}) at the same temperature where the maximum specific heat ($c_{v}^{max}$) occurs. These two characteristics hint at the possibility of a first-order phase transition. However, to definitively characterize a first-order transition from simulations, it is imperative to observe that the transition persists in the thermodynamic limit ($L\rightarrow \infty$). Regarding the specific heat \cite{privman1990finite}, the maximum value exhibits a proportional relationship with the volume of the system dimension, expressed as $c_{v}^{max} \propto L^{d}$.  Therefore, plotting  $c_{v}^{max}$ versus $L^{2}$ for various values of $q$ can reveal if they present a linear dependence. As shown in Fig \ref{Cvmax}, for $q = 1$, no finite-size effect is observed. On the other hand, for q=3 and higher, the specific heat peak increases with the size of the system. However, getting the precise maximum value of the peak becomes increasingly challenging due to fluctuations.

A more effective approach for discerning a first-order phase transition involves computing the histogram of the energy distribution \cite{ferrenberg1988new}. This is achieved through extended runs within the temperature region surrounding the energy discontinuity for different lattice sizes. In a first-order transition, the theory predicts a double-peak structure \cite{binder1987theory}, each peak representing a distinct state of the system: one ordered, the other disordered. The transition temperature is defined when two peaks with equal height emerge in the histogram, indicating equiprobability between the two states. For any $q<3$, no double-peak structure was identified for any lattice size $L$, leading to the conclusion that the only phase transition for the system is of the BKT type.  However, for $q=3$, a subtle double-peak histogram was observed, particularly for $L=64$, with a transition temperature difference of less than 1\% compared to the energy discontinuity and the specific heat peak temperature. As $L$ increases, the double-peak structure becomes more apparent as shown in Fig. \ref{q3Histogram}, because the free energy barrier between the two states becomes higher. The double-peak structure becomes more pronounced with increasing $L$ (see Fig. \ref{q3Histogram}), reflecting a growing free energy barrier between the two states and, again, the transition temperature differs by less than 1\% from the energy discontinuity and specific heat peak temperatures.

To confirm the first-order transition, we will employ the Lee-Kosterlitz criterion \cite{lee1991finite}. According to this criterion, determining a first-order transition involves calculating the free energy barrier $\triangle F (L)$ across different system sizes $L$. So, the criterion implies that $\triangle F (L)$ should be independent of $L$, and at a first-order phase transition, it is expected to be an increasing function of $L$. For $q=3$, the double-peak structure shows up faintly for $L=64$. For higher values of $q$, the double-peak structure {\color{black} manifests itself} much earlier, for example, in $q=4$ we already can observe it for $L=32$. This inherent size dependence poses challenges in applying the Lee-Kosterlitz criterion to small values of $q$. A similar problem was observed by Lee and Kosterlitz \cite{lee1991finite} in their paper concerning the {\color{black}Potts} model with small values of the generalization parameter. They attributed this issue to the weakness of the transition. However, for $q=3,5,6$, we observe the anticipated behavior indicative of a first-order phase transition, as depicted in Fig. \ref{LeeKosterlitz}. Despite computational limitations preventing the {\color{black} application} of the Lee-Kosterlitz criterion, the observed increase in the free energy barrier supports the assertion that the critical parameter is $q=3$. In Table \ref{TabCrit}, we show the values of the transition temperature utilizing the Binder Fourth Cumulant (see Appendix \ref{appendix:Binder}). This method reveals another characteristic of a first-order transition: a jump to a negative value precedes the attainment of the value $U_{L} = 0.5$ for an ordered system \cite{tsai1998fourth}.

In the range from $q=3$ to $q=5$, the critical temperature obtained through finite-size scaling of the susceptibility and the Helicity (see Appendix \ref{appendix:FFSMagnetic} and Appendix \ref{appendix:Helicity} respectively) consistently reveals transition temperatures higher than the critical temperature obtained. However, for $q=6$ and higher, the critical temperature and the transition temperature obtained from the Binder Fourth Cumulant and the susceptibility converge. Consequently, the determination of the critical temperature becomes challenging based on the finite-size scaling of the susceptibility and Binder Fourth Cumulant. This observation potentially suggests the absence of the BKT transition for this generalized parameter. 

\begin{figure}
    \centering
    \includegraphics[width=8cm]{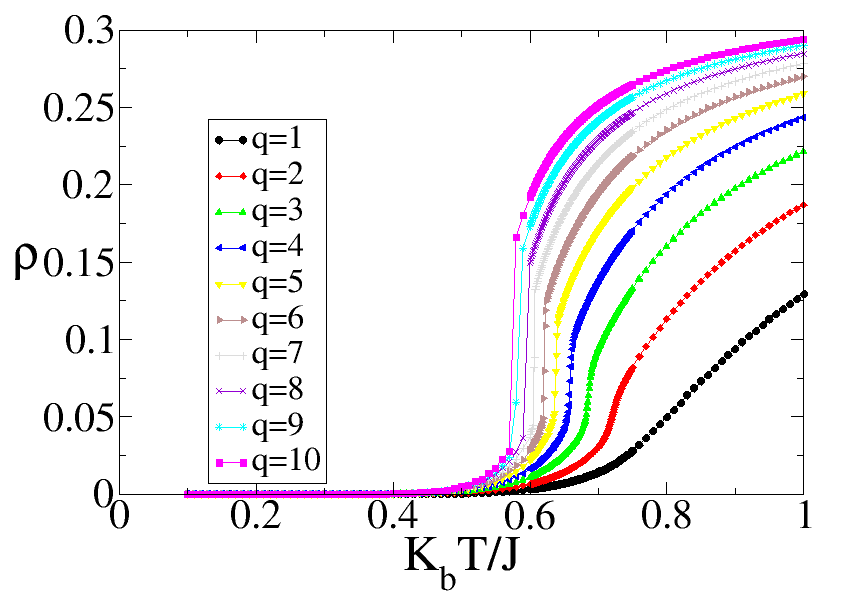}
    \caption{Vortex Density as a function of temperature for different q values, and $L=96$. We can observe a discontinuity being formed with the increase of q.}
    \label{TxVortice}
\end{figure}

To determine the critical temperature, the only method that was possible to determine a critical temperature was through the finite-size scaling (FSS) of the helicity modulus. This is because the system exhibits vortices and the population of these topological objects increases rapidly as the generalization parameter $q$ increases, as illustrated in \ref{TxVortice}.

The vortex density exhibits a discontinuity for sufficiently large $q$ and such discontinuity occurs at the same temperature as the energy and where the specific heat presents a peak, indicating a correlation between these events. To illustrate this correlation more effectively, we calculated the vortex density difference as an order parameter. Essentially, when the system transitions between states, the vortex density difference jumps from 0 to a non-zero value. Such behavior allows us to define a way to get the critical generalization parameter of the model, simply by calculating the mean vortex density difference for different q at the region of temperature where supposedly the discontinuity should appear. The first value where the mean vortex density difference is appreciable is given by $q=3$, which agrees with the result obtained by the histogram and Binder Cumulant.
%The vortex density exhibits a discontinuity with the increase of $q$ at the same temperature as the energy and where the specific heat presents a peak, indicating a correlation between these events. To illustrate this correlation more effectively, we calculated the vortex density difference as an order parameter. Essentially, when the system transitions between states, the vortex density difference jumps from 0 to a non-zero value. To better demonstrate we took the vortex density difference as an order parameter, that is, when the system changes from one state to another, the vortex density difference jumps from 0 to a specified value. That allowed us to define a way to obtain the critical generalization parameter of the model, simply by calculating the mean vortex density difference for different $q$ at the region of temperature where supposedly the discontinuity should appear. The first value where the mean vortex density difference was appreciable, was for $q=3$, agreeing with the result obtained by the histogram and Binder Cumulant. 
%Essentially, when the system transitions between states, the vortex density difference jumps from 0 to a non-zero value. We observed the first appreciable value for $q=3$, aligning with the result obtained from the histogram and Binder Cumulant.

\begin{figure*}
\begin{minipage}{.32\textwidth}
    \subfloat["Ordered Phase"]{\includegraphics[width=\textwidth]{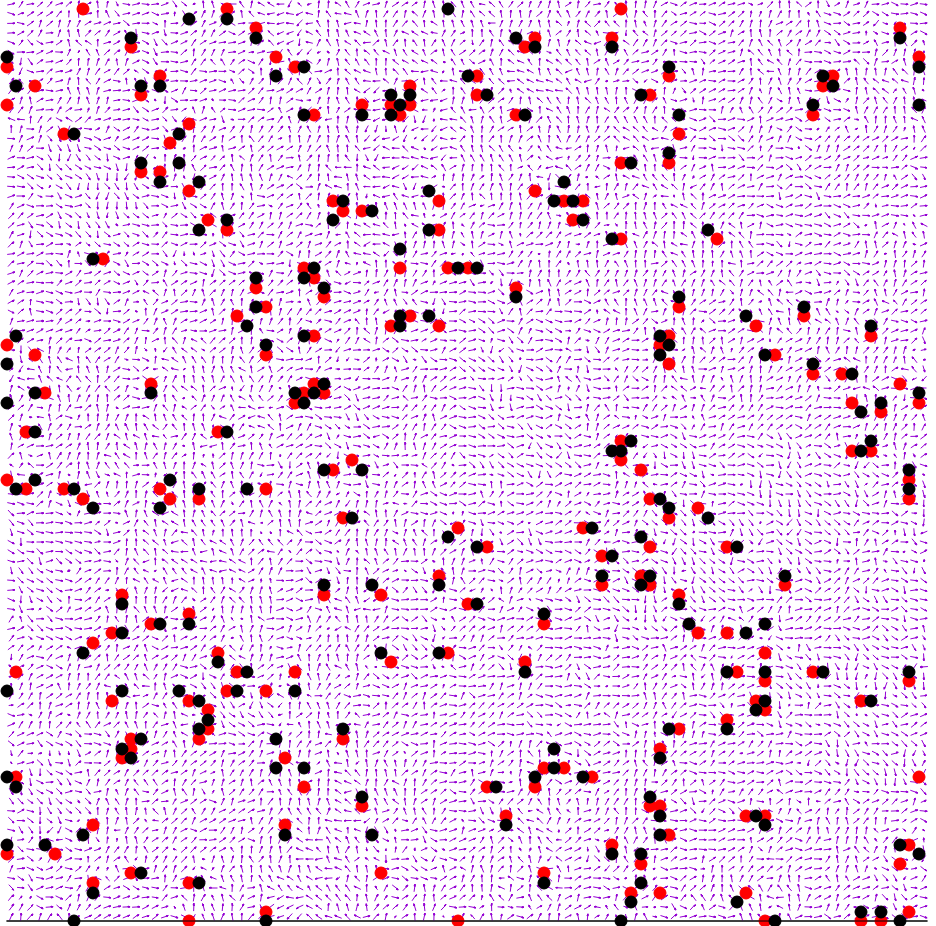}\label{VortexLow}}
\end{minipage}
\hfill    
\begin{minipage}{.32\textwidth}
    \subfloat[q=6 Histogram]{\includegraphics[width=\textwidth]{Graficos/q6Hist.png}\label{Histq62}}
\end{minipage}
\hfill    
\begin{minipage}{.32\textwidth}
    \subfloat["Disordered Phase"]{\includegraphics[width=\textwidth]{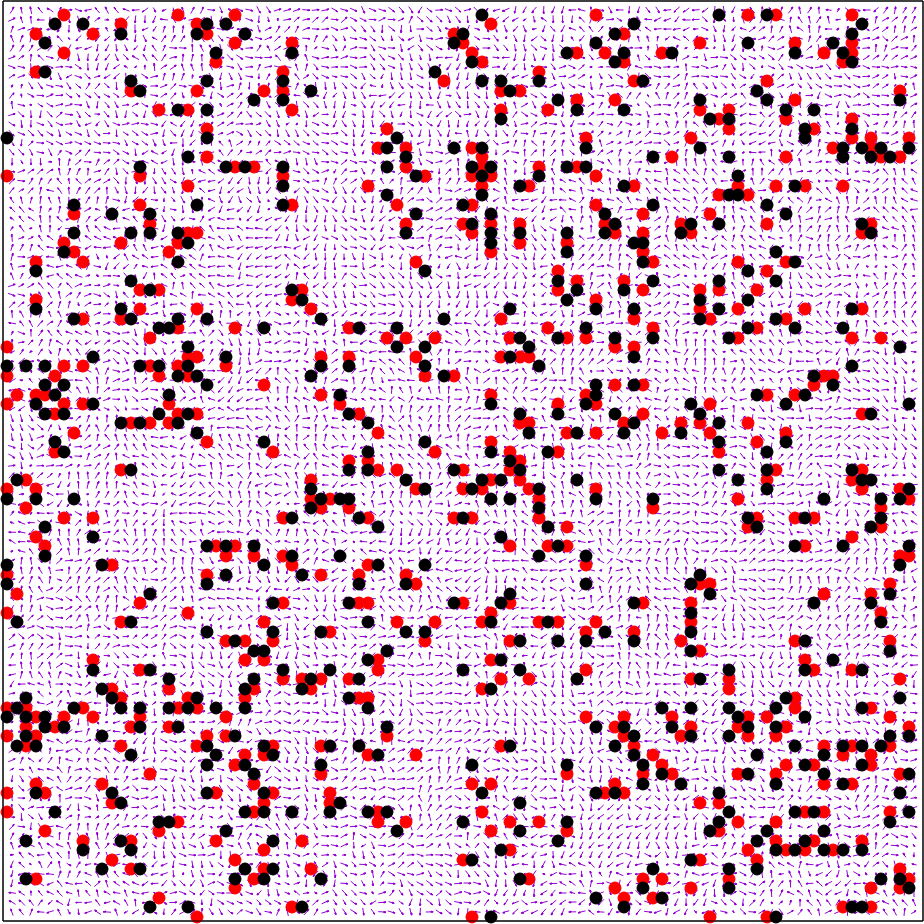}\label{VortexHigh}}
\end{minipage}
    \caption{Two distinct configurations were observed for $q=6$ at the system's transition temperature, $T_{c}=0.6212$, with a lattice size of $L=96$. The red points denote vortices, while the black points represent antivortices. The state labeled "Disordered" with higher energy, exhibits a significantly larger population of vortices compared to the "Ordered" state with lower energy.}\label{q6Histgram}
\end{figure*}

\subsection{\label{sec:level2}Gibbs Measures and First-Order Transitions}

The Gibbs Measures method is a powerful tool for studying first-order transitions. For every Hamiltonian $H^{\Lambda}_{\epsilon}(\phi,\theta)$, there exists a Gibbs measure $\mu^{\Lambda}(d\phi, d\theta)$.  If this measure is not unique, the system exhibits a first-order transition. A solid background on Gibbs measures and phase transitions can be found in \cite{georgii2011gibbs,georgii2001random}.  One of the early contributions using Gibbs Measures was made by Dobrushin and Shlosman \cite{dobrushin1975absence}, where they established a theorem stating that every state of Gibbs on a two-dimensional lattice with a continuous, inferiorly bounded, translationally invariant and rapidly decreasing potential is always an invariant measure under the action of a symmetry group G.

While these conditions may appear restrictive, they are not impossible to satisfy. Indeed, the Heisenberg model meets these conditions, and this theorem essentially extends the Mermin-Wagner theorem, wherein invariance to the potential means the absence of continuous symmetry breaking in two-dimensional systems. The possibility of phase transitions in two-dimensional models with continuous symmetries was further explored by Shlosman in another paper\cite{shlosman1980phase}, utilizing two powerful methods: The chessboard estimate and the reflection positivity \cite{frohlich2004phase,frohlich1978phase,shlosman1986method,biskup2009reflection}. Shlosman demonstrated that obtaining two phases for isotropic short-range interaction models with continuous symmetry is possible, provided the temperature is low enough. However, all phases remain invariant under an action called $G$ in the configuration space, meaning that there is {\color{black} certainty} a first-order phase transition without breaking symmetry. As a {\color{black} another example}, Van Enter and Shlosman\cite{van2002first} proved this for the Domany model\cite{domany1984first}.

Our focus here lies on the results obtained by Van Enter et al.\cite{van2006first} for the generalized XY model. They employed a square-ditch approximation, where only spins within the same ditch can interact. The present implies that a small interval $[\pi/2 - \epsilon, \pi/2 + \epsilon]$, is defined for the polar angle $\theta$, where the ditch parameter $\epsilon$ is a function of the $q$-parameter, i.e. $\epsilon(q)$. In addition, $\epsilon(q)$ must decay at least with $1/\sqrt{q}$. This approach implies that two spins interact only when they are in the same ditch. The aforementioned allows them, by utilizing the reflection positivity and the chessboard estimate, to obtain an estimation of the partition function $Z^{\Lambda}$, which is given by,

\begin{equation} \label{eq:zl}
     Z^{\Lambda} \geq (C_{1} \epsilon \exp (2 C_{2} \beta ) )^{|\Lambda|} \quad .
 \end{equation}
where $\beta$ is the temperature, and they assume that $C_{1} = C_{2} \approx \cos( \pi/20) \approx 1 $. To demonstrate the possibility of a first-order phase transition, they devised another partition function. This function considers a configuration divided into two regions: an "ordered" region where the central site and its neighbors all reside within the ditch and a "disordered" region where the central spin and its neighbors are not in the same ditch. They integrated over all configurations compatible with this contour, resulting in a partition function with a universal contour,$Z^{\Lambda}_{uc}$  given by:

 \begin{equation} \label{eq:zluc}
    Z^{\Lambda}_{uc} \leq ( (2 \epsilon )^{3/4} \exp(\beta)  )^{| \Lambda |} \quad .
\end{equation}

From the estimates of the partition functions, it is possible to demonstrate, using theorem 1 in the paper by Van Enter et al. \cite{van2002first}, the coexistence of two Gibbs measures. This is achieved by considering

\begin{equation}
    \frac{Z^{\Lambda}_{uc}}{Z^{\Lambda}} \leq \epsilon^{\frac{|\Lambda |}{(4 + C_{3})}}
\end{equation}
In essence, if $\epsilon$ is less than 1, $Z^{\Lambda}$ dominates the fraction, resulting in a suppression of the contours between the ordered and disordered sites, leading to a completely disordered system. If $\epsilon$ exceeds 1, the partition function $Z^{\Lambda}_{uc}$ dominates this fraction and we have a "ordered" system. Thus, it is possible to exist a temperature region where the two partition functions are comparable, implying the non-uniqueness of the Gibbs measures. %, as will be shown shortly.$

In the work of Van Enter et al. \cite{van2006first}, they did not estimate the critical temperature or the critical parameter q where the first-order transition appears. Here, we define a function for $\epsilon$ that preserves the system's interactions on average as a function of the parameter $q$. This function is given by a combination of gamma functions as follows: (for a detailed derivation, see Appendix \ref{appendix:epsilon}) %What we did is define a function for $\epsilon$, constructed by gamma functions as a function of parameter q, that preserves the mean value of interactions in the system. This function is given by:
\begin{equation}
 \epsilon(q) =  \frac{\sqrt{\pi}}{2}  \left ( \frac{2^{-q} \sqrt{\pi} \Gamma[\frac{1}{2} + q] }{\Gamma\left[ \frac{1+q}{2} \right]^{2}} \right )^{-1} \frac{\Gamma \left[ \frac{1+q}{2}\right ] }{\Gamma \left [ 1 + \frac{q}{2} \right ]  }   
\end{equation}

Applying the conditions for a first-order phase transition ($\epsilon \leq 1$), we obtained the critical parameter $q_{c} \geq 2.28851$. {\color{black} Note that $q$ is a measure for
the narrowness of the ditch which can vary continuously. However,
since we are considering only integer values of $q$}, the smallest integer critical parameter would be $q_{c} = 3$, which coincides with the value obtained through simulations. The previous equation also allows us to estimate the transition temperature for any $q$ value. With this equation in hand, we can again divide eq.(\ref{eq:zluc}) by eq.(\ref{eq:zl}) and, through a simple algebraic manipulation, as follows, we get:

\begin{table*}
\caption{\label{TabCrit}Critical and transition temperature for different values of q by the methods discussed.}
\begin{ruledtabular}
\begin{tabular}{cccccccc}
 \multicolumn{1}{c}
 {\color{black} q} & \color{black} FSS $\chi$ & \color{black} FSS $\Upsilon$& \color{black} BKT $U_L$ & \color{black} First-Order $U_L$ & \color{black} Histogram & \color{black} Gibbs Measures\\ \hline
 \textbf{1} & $0.699\pm 0.001$    & $0.692 \pm 0.005$         & $0.710 \pm 0.001$    & $------$                      & $------$             & $-----$                    \\  
 \textbf{2} & $0.679\pm 0.003$    & $0.662 \pm 0.005$         & $0.688 \pm 0.003$    & $------$                      & $------$             & $-----$                    \\
 \textbf{3} & $0.655\pm 0.001$    & $0.638 \pm 0.005$         & $0.657 \pm 0.005$    & $0.682 \pm 0.005$             & $0.685 \pm 0.001$     & $0.68 \pm 0.05$           \\
 \textbf{4} & $0.634\pm 0.001$    & $0.621 \pm 0.005$         & $0.633 \pm 0.003$    & $0.662 \pm 0.003$             & $0.657 \pm 0.001$     & $0.65 \pm 0.05$           \\
 \textbf{5} & $0.616\pm 0.001$    & $0.608 \pm 0.005$         & $0.613 \pm 0.004$    & $0.640 \pm 0.004$             & $0.637 \pm 0.001$     & $0.64 \pm 0.05$           \\ 
 \textbf{6} & $------$            & $0.599 \pm 0.005$         & $------$             & $0.623 \pm 0.002$             & $0.621 \pm 0.001$     & $0.62 \pm 0.04$           \\ 
 \textbf{7} & $------$            & $0.586 \pm 0.005$         & $------$             & $0.608 \pm 0.001$             & $0.607 \pm 0.001$     & $0.61 \pm 0.04$           \\
\end{tabular}
\end{ruledtabular}
\end{table*}

\begin{equation}
    \frac{Z^{\Lambda}_{uc}}{Z^{\Lambda}} \leq \frac{|(2 \epsilon)^{\frac{3}{4}} \exp(\beta)|^{|\Lambda |}}{|C_{1}\epsilon \exp(2 C_{2} \beta )|^{|\Lambda |}} \leq 1
\end{equation}

%because $ Z^{\Lambda} \leq 1 $ e $Z^{\Lambda}_{uc} \leq |(2 \epsilon)^{\frac{3}{4}}exp(\beta)|^{|\Lambda |} $. So we have,
%

\begin{equation}
    \exp[\beta(1-2C_{2}) - \frac{1}{4}ln(\epsilon) + \frac{3}{4}ln(2) - ln(C_{1})] \leq 1
\end{equation}
\begin{equation}
    \beta(1-2C_{2}) - \frac{1}{4}ln(\epsilon(q)) + \frac{3}{4}ln(2) - ln(C_{1}) \leq 0
\end{equation}
Thus, we can estimate the critical temperature at which the first-order phase transition occurs by letting $C_{1}$ and $C_{2}$ be free parameters, expecting them to be lower than $\cos(45 ^{\circ})$ in agreement with the analytical results of Van Enter et al. The results are shown in Table \ref{TabCrit} and are consistent with the transition temperature obtained by histogram and Binder Fourth Cumulant. The angles obtained for both $C_{1}$ and $C_{2}$ were close to $\sim 40^{\circ}$ regardless of the parameter $q$, a value lower than that was obtained in the simulations considering the entire lattice, which was approximately $\sim 52 ^{\circ}$ in the transition temperature. Notably, the value of $q$ doesn't seem to affect the azimuthal angle value.

\section{conclusion}

In this study, we have employed Monte Carlo simulations and analytical analysis to investigate the 2D generalized XY model. Our findings reveal not only characteristics of a first-order phase transition for $q \geq 3$, but also the presence of three distinct regions characterized by the existence of phase transitions. In the first region ($q \le 3$), a BKT regime is observed where the energy and the vortex density vary continuously, consistent with the usual XY model. In the second region ($3 \le q < 6$), a coexistence of first-order transition and BKT transition occurs, but at different temperatures, with the first-order transition temperature always higher than the BKT transition. In the third region ($q \ge 6$), the first-order transition persists, but the analyses for the critical temperature of the BKT transition do not provide conclusive results. This suggests a possible change in the universality of the system and the potential extinction of the BKT phase transition.

Another important result is the impact of the vortex density on the phase transitions of the system. To illustrate, consider the histogram for $q=6$ at the transition temperature $T_{c}=0.6212$ and $L=96$ (figure \ref{Histq62}). The two different phases of the system are represented by figures \ref{VortexHigh} and \ref{VortexLow}. Figure \ref{VortexLow} represents an "Ordered Phase" with a lower vortex density, while figure \ref{VortexHigh} is a "Disordered Phase" with a higher vortex density. {\color{black} The narrowness of the ditch causes an entropy jump \cite{georgii2001entropy}, leading to the proliferation of vortices in the system. This is identified as the cause of the first-order phase transition.}
To connect this result with the findings of Van Enter et al., consider the "disordered" state, where two spins are in different ditches. The weak interaction between the spins increases the energy of the system, leading to the proliferation of vortices as shown in \ref{VortexHigh}. Conversely, when the spins belong to the same ditch, the energy decreases rapidly, resulting in a decrease in the number of vortices. Therefore, the first-order transition from an "ordered" state to a "disordered state" is caused by a proliferation of vortices in the system. It is essential to note that the quotation marks around "ordered" and "disordered" emphasize the absence of real order in the system. One phase is merely more ordered than the other, and there is no actual order in the system. Additionally, the magnetization remains null in agreement with the Theorem of Mermin-Wagner and {\color{black}Van Enter-Shlosman rigorous proof}, making it a {\color{black} disordered-disordered \cite{enter2005provable} } first-order transition without symmetry breaking. 

In summary, our study offers a comprehensive understanding of the properties and phase transition behavior of the generalized XY model. The results obtained confirms the occurrence of first-order phase transitions and elucidate their underlying mechanism. The identified classes of phase transitions align with the analytical results, and the mechanism of this new phase transition is attributed to the proliferation of vortices in the system. The next step is to explore whether other models exhibiting first-order phase transitions, such as the Domany model, can be explained similarly.

\begin{acknowledgments}
This study was financed in part by the Coordenação de Aperfeiçoamento de Pessoal de Nível Superior (CAPES) - Brazil - Finance Code 001. R.J.C.L. acknowledges A. Tirelli for the valuable discussions and M. Capone for all the support in developing and finalizing this work. R.J.C.L. acknowledges the financial support of the National Recovery and Resilience Plan PNRR MUR, PE0000023-NQSTI, financed by the European Union - Next Generation EU and the MUR Italian National Centre for High-Performance Computing, Big Data and Quantum Computing, grant number CN00000013, as well as the financial support from Simon's Foundation. %grant no. 319270FY19 

\end{acknowledgments}

%\nocite{*}
\bibliographystyle{unsrt}
\bibliography{bibliography.bib}% Produces the bibliography via BibTeX.

%\newpage

\appendix

\color{black}
\section{Obtaining the expression for $\epsilon(q)$}
\label{appendix:epsilon}

\begin{figure}[!ht]
    \centering
    \includegraphics[width=1.00\linewidth]{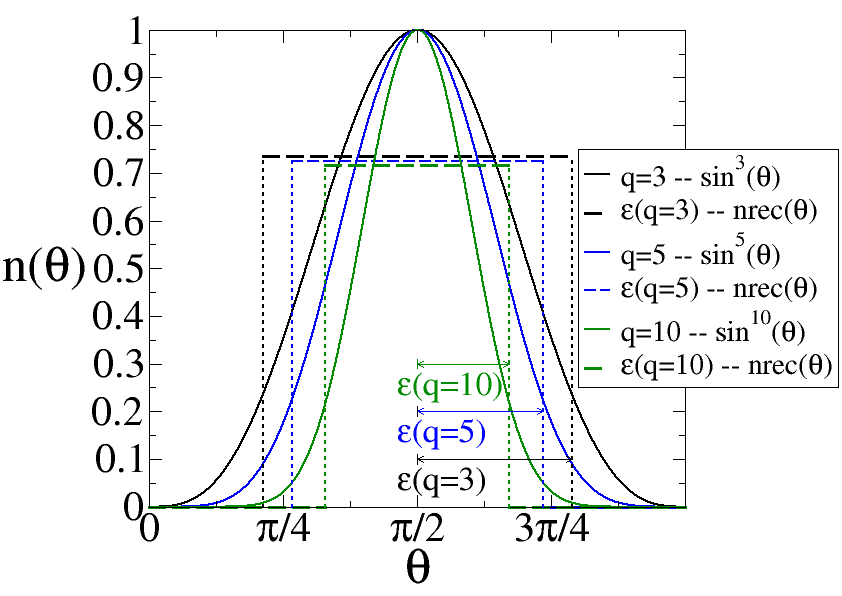}
    {\color{black} \caption{Graph showing the original $n(\theta)$ curve (solid lines) and the respective boxcar $nrec(\theta)$ curves (dashed lines) for different $q$ values ($q=3,5,10$).}}
    \label{fig:figepsilon}
\end{figure}

The following steps are considered to get a q-dependent expression for $\epsilon$. First, $n(\theta)$ is replaced in the original Hamiltonian by a boxcar function, $nrec(\theta)$, as shown in the fig.(\ref{fig:figepsilon}). The boxcar function is defined in terms of two Heaviside functions, as follows:
\begin{equation}
    nrec(\theta) = A\Big( H\Big[\theta - \Big(\frac{\pi}{2}-\epsilon(q)\Big)\Big] - H\Big[\theta - \Big(\frac{\pi}{2}+\epsilon(q)\Big)\Big] \Big)
\end{equation}
where $A$ is the height and $H[x]$ is a Heaviside function, which is equal $0$ if $x<0$ and equals to $1$ otherwise.

After the previous substitution, in order to preserve the average interactions of the system, we require that both curves, $n(\theta)$ and $nrec(\theta)$, have the same area as well as the height of $nrec(\theta)$ is equal to the average height of $n(\theta)$. Then, the equality between the areas gives:
\begin{equation*}
    \mathcal{A}[nrec(\theta)] = \mathcal{A}[n(\theta)]
\end{equation*}

\begin{equation}\label{eq:area}
    A\Big(2\epsilon(q)\Big) = \int_0^\pi \mathrm{d}\theta \sin^q(\theta) = \frac{\sqrt{\pi } \Gamma \left(\frac{q+1}{2}\right)}{\Gamma \left(\frac{q}{2}+1\right)}
\end{equation}
and, the equality of the height of $nrec(\theta)$ with the average height of $n(\theta)$ gives:
\begin{equation*}
    A = \int_0^\pi \mathrm{d}\theta \left( \frac{\sqrt{\pi } \Gamma \left(\frac{q+1}{2}\right)}{\Gamma \left(\frac{q}{2}+1\right)} \right)^{-1} \sin^{2q}(\theta)
\end{equation*}

\begin{equation} \label{eq:height}
    A = \frac{2^{-q} \sqrt{\pi } \Gamma \left(q+\frac{1}{2}\right)}{\Gamma \left(\frac{q+1}{2}\right)^2}
\end{equation}

 Substituting Eq.(\ref{eq:height}) in Eq.(\ref{eq:area}), we get the equation for $\epsilon(q)$, given by:
\begin{equation*}
    2\epsilon(q) \frac{2^{-q} \sqrt{\pi } \Gamma \left(q+\frac{1}{2}\right)}{\Gamma \left(\frac{q+1}{2}\right)^2} = \frac{\sqrt{\pi } \Gamma \left(\frac{q+1}{2}\right)}{\Gamma \left(\frac{q}{2}+1\right)}
\end{equation*}

\begin{equation} \label{eq:epsq}
    \epsilon(q) = \frac{\sqrt{\pi}}{2} \left( \frac{2^{-q} \sqrt{\pi } \Gamma \left(q+\frac{1}{2}\right)}{\Gamma \left(\frac{q+1}{2}\right)^2} \right)^{-1} \frac{\Gamma \left(\frac{q+1}{2}\right)}{\Gamma \left(\frac{q}{2}+1\right)}
\end{equation}

An essential property of the ditch width, $\epsilon(q)$, as stated by \cite{van2006first}, is that for large values of $q$, the spins only interact effectively for very narrow ditches around $\theta=\pi/2$, whose width is of the order of $\mathcal{O}(1/\sqrt{q})$. This can be easily verified from Eq.(\ref{eq:epsq}) by applying Stirling's approximation to the gamma functions assuming large values of q, from which we get:
\begin{equation}
    \epsilon(q) \xrightarrow[q \to \infty]{} \frac{\sqrt{\pi}}{\sqrt{q}}
\end{equation}
as expected.

\color{black}

\section{Binder's Fourth Cumulant}
\label{appendix:Binder}

\begin{figure*}
\begin{minipage}{.32\textwidth}
    \subfloat[q=3 Binder Cumulant]{\includegraphics[width=\textwidth]{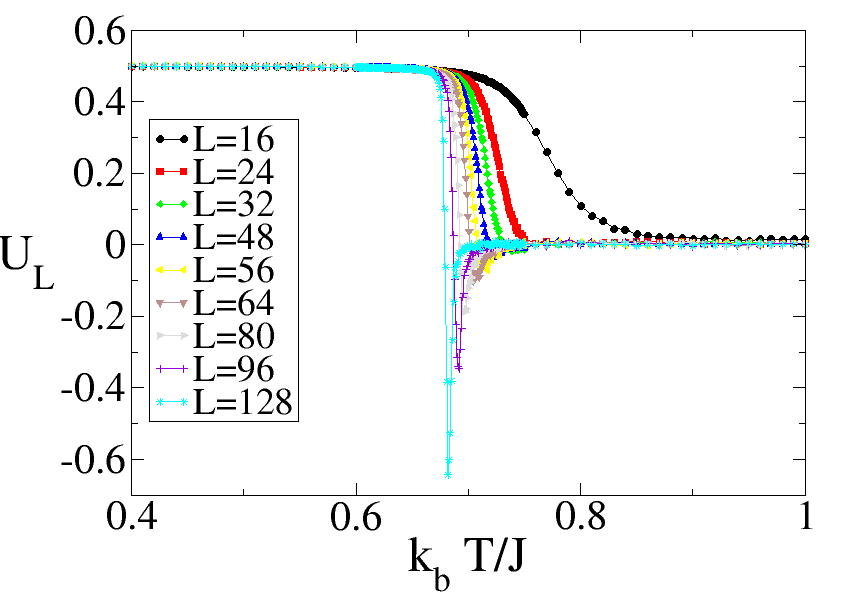}\label{BinderCumulantq3}}
\end{minipage}
\hfill    
\begin{minipage}{.32\textwidth}
    \subfloat[q=3 Binder Cumulant]{\includegraphics[width=\textwidth]{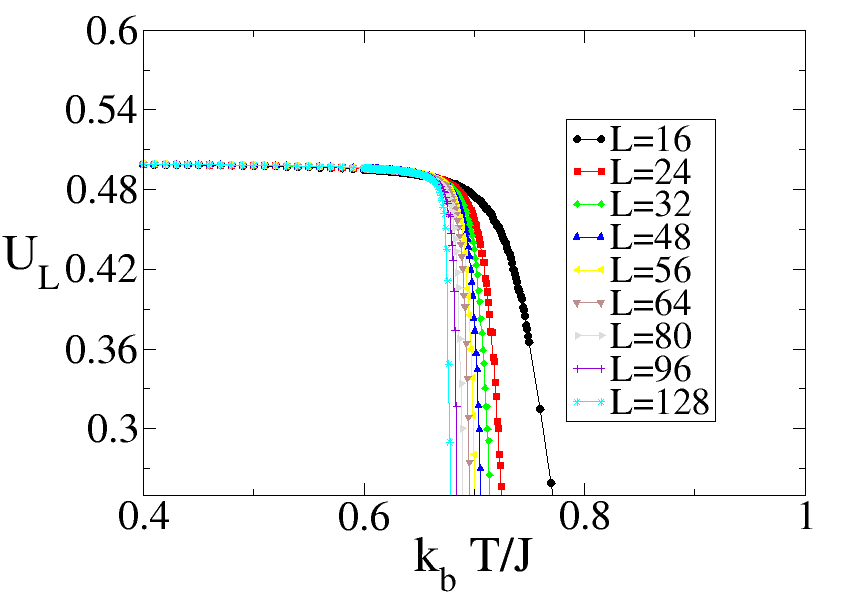}\label{fig:sub_a2}}
\end{minipage}
\hfill 
\begin{minipage}{.32\textwidth}
    \subfloat[q=6 Binder Cumulant]{\includegraphics[width=\textwidth]{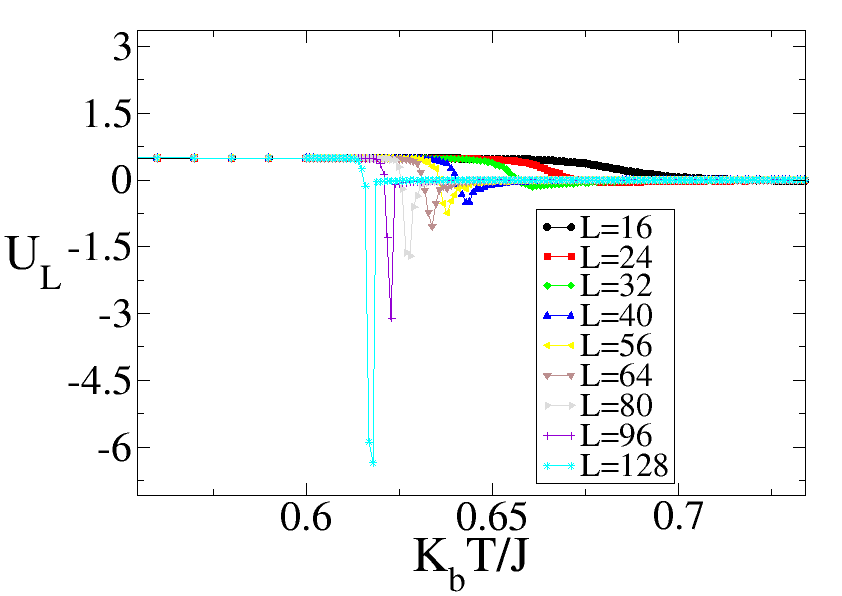}\label{TxBq6}}
\end{minipage}
\hfill
\begin{minipage}{.32\textwidth}
    \subfloat[\centering q=3 FSS]{\includegraphics[width=\textwidth]{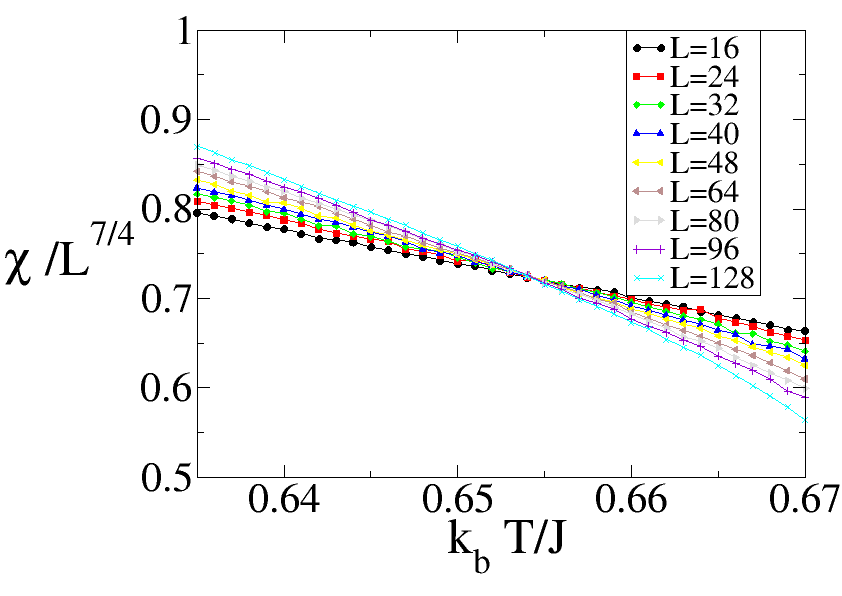}\label{FSSMagneticSusceptibility}}
\end{minipage}
\hfill 
\begin{minipage}{.32\textwidth}
    \subfloat[\centering q=3 FSS]{\includegraphics[width=\textwidth]{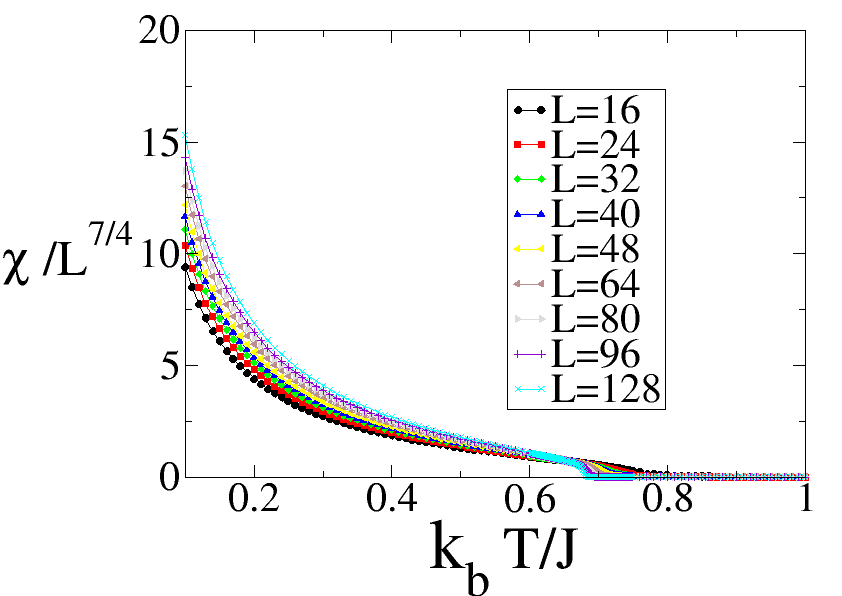}\label{FSSMagneticSusceptibility}}
\end{minipage}
\hfill 
\begin{minipage}{.32\textwidth}
    \subfloat[\centering q=6 FSS]{\includegraphics[width=\textwidth]{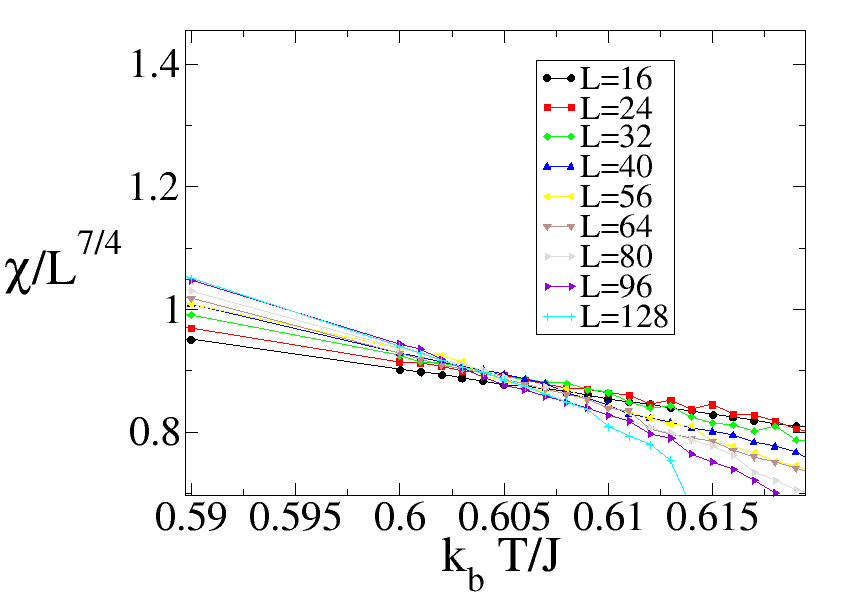}\label{FSSMagneticSusceptibility}}
\end{minipage}
\hfill 
\caption{(a),(b) is Binder Cumulant of magnetization for $q=3$ for several lattice sizes. Both crossing points for the BKT transition and the first-order phase transition region can be observed. (c) Binder Cumulant of magnetization for $q=6$ across various lattice sizes. The crossing point for the BKT transition cannot be observed anymore, but the first-order phase transition region is still visible. (c),(d) FSS of magnetic susceptibility for $q=3$ and $q=6$. For $q=3$ it's possible to determine one crossing point, while for $q=6$ there's a region of crossing points.}\label{fig:fig}
\end{figure*}

An initial estimation of the critical temperature can be derived from the Binder's Fourth Cumulant defined as:

\begin{equation}
U_L = 1 - \frac{\left \langle \left (M_x^2 + M_y^2 \right )^2 \right \rangle}{2\left \langle M_x^2 + M_y^2 \right \rangle}
\end{equation}
where $M_x$ and $M_y$ represent the in-plane magnetization components. For any system size, this quantity reaches $0.5$ in the low-temperature phase $(T \ll T_c)$ and tends towards zero in the high-temperature phase $(T \gg T_c)$. When measured at $T_c$, its value is approximately independent of the system size. Therefore, $T_c$ can be estimated by plotting $U_L$ vs $T$ for different system sizes and identifying the point of intersection. In a first-order transition, the Cumulant assumes negative values \cite{tsai1998fourth,lee1991finite}.

As an example, consider the parameter $q=3$ as in Figure \ref{BinderCumulantq3}. We can observe the two transitions, a region where the cumulant of {\color{black} Binder} assumes negative values with a minimum at $T=0.68(2)$ for $L=128$ represents a first-order transition, and a crossing point in $T=0.65(7)$ supposedly represents a BKT phase transition.

From $q=1$ to $q=5$ we can observe the BKT transition at the point where the curves intersect, however, for $q=6$ and higher the curves overlap in more than one point, so it is not possible anymore to estimate the critical temperature this way. It is a reliable technique, but different methods for determining the $T_c$ were considered.

\section{finite size scaling of magnetic susceptibility}
\label{appendix:FFSMagnetic}

A more accurate estimate for the critical temperature for a BKT transition involves the finite-size scaling analysis of the in-plane susceptibility $\chi$. To obtain the susceptibility of a component $\alpha$, the magnetization fluctuations are calculated as:

\begin{equation}
\chi^{\alpha \alpha} = \frac{1}{N k_{b} T} \left( \langle M_{\alpha}^{2} \rangle - \langle M_{\alpha} \rangle^{2} \right)
\end{equation}
where $M_{\alpha}$ represents the value of the magnetization obtained by summing over the component $\alpha$ for all spins of the lattice. The in-plane susceptibility is the average of the planar component susceptibilities:

\begin{equation}
\label{eq:Inplanesusceptbility}
\chi = \frac{1}{2} \left( \chi^{x x} + \chi^{y y} \right)
\end{equation}

The finite-size scaling of the in-plane susceptibility is derived assuming that a power-law scaling of the susceptibility \cite{kosterlitz1973ordering} holds near and below $T_c$ for any value of q,

\begin{equation}
\chi \propto L^{2 - \eta}
\end{equation}
the exponent $\eta$ describes the long-distance behavior of in-plane correlations below $T_c$. For the XY model, the critical temperature is reached when $\eta = 1/4$. Initially, we assume that $\eta$ is valid for any value of $q$, and the assumption is tested by the quality of the produced scaling. Thus, $\chi / L^{7/4}$ is plotted against $t$ for various system sizes. The common crossing point of the curves yields the critical temperature. For $q=1$, the expected result for the critical temperature of $T_{c}=0.69(9)$ is obtained. With an increase in the parameter $q$, the critical temperature decreases. For example, in Figure \ref{FSSMagneticSusceptibility}, the critical temperature obtained was $t=0.65(5)$, and for $q=6$, where determining the critical temperature by this method becomes challenging, a region of crossing points is observed, as depicted in Figure \ref{FSSMagneticSusceptibility}. This behavior persists for higher values of $q$. The interpretation drawn is that the assumption of the value of the scale parameter is no longer valid, indicating a change in the order of the phase transition.

\begin{figure*}
\begin{minipage}{.32\textwidth}
    \subfloat[\centering $q=3$]{\includegraphics[width=\textwidth]{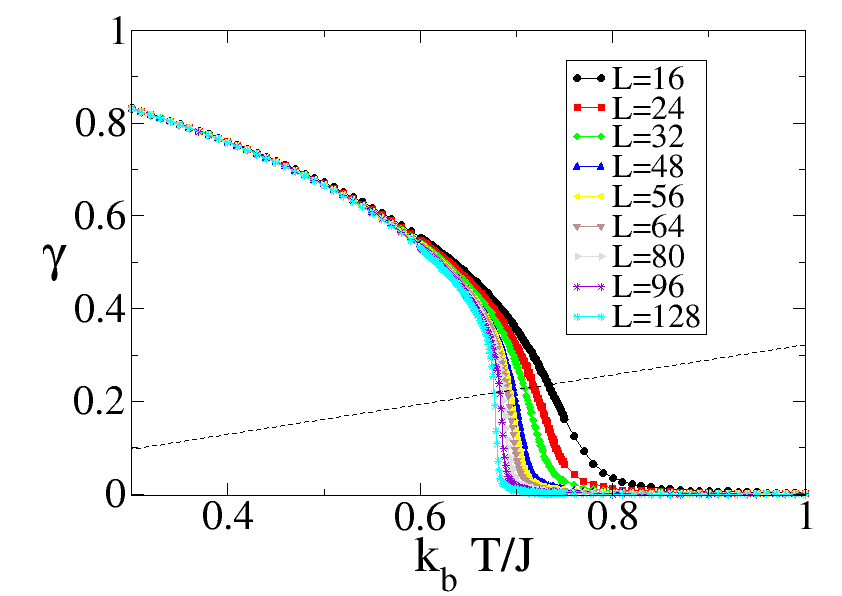}\label{Helicity}}
\end{minipage}
\hfill    
\begin{minipage}{.32\textwidth}
    \subfloat[\centering $q=6$]{\includegraphics[width=\textwidth]{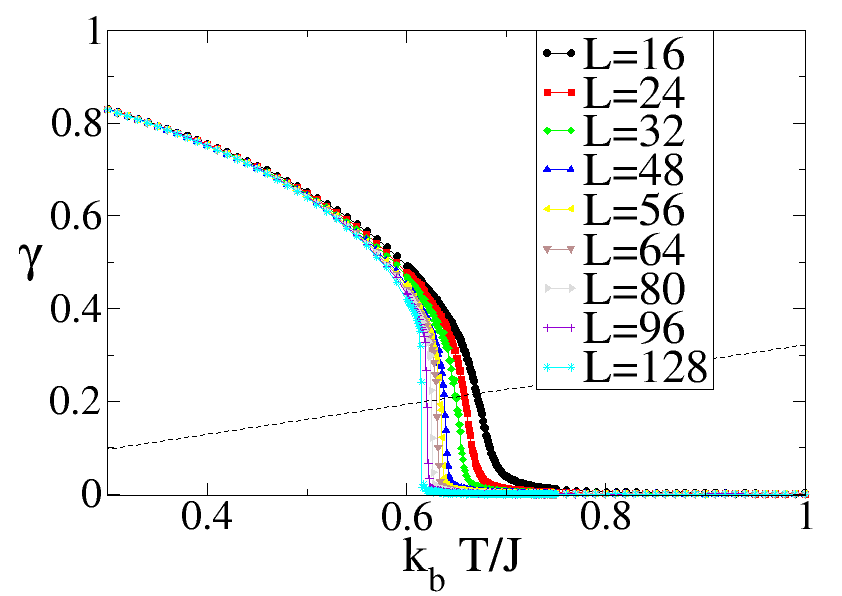}\label{Helicity}}
\end{minipage}
\hfill    
\begin{minipage}{.32\textwidth}
    \subfloat[q=6 Binder Cumulant]{\includegraphics[width=\textwidth]{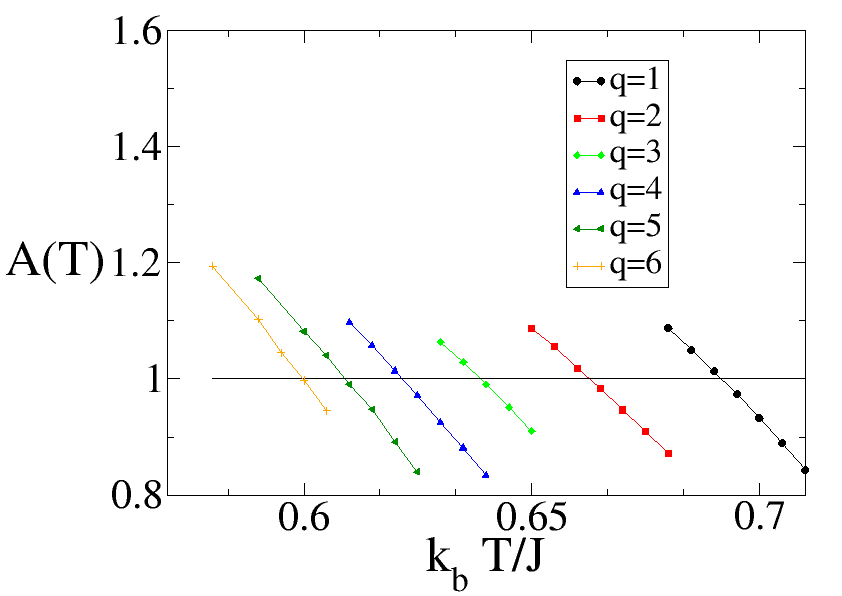}\label{AT}}
\end{minipage}
    \caption{Effects of finite-size scaling of helicity modulus for two different values of $q$ is observed in (a) and (b). With the increase of $q$, the helicity jump becomes inclined and starts at lower temperatures. (c) show the FSS of the helicity modulus, from $q=1$ to $q=6$, the point where the two lines cross is the expected critical temperature.}
\end{figure*}

\section{Helicity Modulus and Finite-Size Scalling}
\label{appendix:Helicity}

The Helicity is a measure of the change in dimensionless free energy due to an infinitesimal spin twist across the system along one coordinate,

\begin{equation}
\Upsilon (T) = \frac{1}{N} \frac{\partial^2 f}{\partial \Delta^2}
\end{equation}
taking an infinitesimal spin twist at the x component for the generalized XY model:

\begin{equation} \label{eq1}
\begin{split}
\Upsilon (T) = & -\frac{1}{2} \left \langle H \right \rangle \\
 & -\frac{J}{N k_b T} \left \langle \left [ \sum_{\left \langle i,j \right \rangle}^{} (\sin\theta_i \sin\theta_j)^q \sin( \phi_i - \phi_j ) \hat{e}_{ij} \cdot \hat{x} \right ]^{2} \right \rangle
\end{split}
\end{equation}

According to the renormalization group (RG), the helicity modulus in an infinite system jumps from zero to a finite value $\frac{2}{\pi}k_B T_c$ at the critical temperature. {\color{black} Although such a result was firstly achieved by an RG
argument, it has also been rigorously obtained for the two-dimensional XY Model by Chayes\cite{CMP}.} Therefore, we can obtain $T_c$ by plotting the helicity as a function of the temperature and locating the intersection with the straight line. Due to the dependence on the size of the lattice, better results can be obtained using finite-size scaling of the system. A useful scaling expression \cite{wysin2005extinction,harada1997universal,capriotti2003reentrant,cuccoli2003quantum} for BKT is given by:

\begin{equation}
\frac{\pi \Upsilon}{2 k_B T} = A(T) \left[ 1 + \frac{1}{2\ln\left(\frac{L}{L_o}\right)} \right]
\end{equation}
where $A(T)$ and $L_o$ are fitting constants. This expression is exact at $T=T_c$, where $A(T_c)=1$. By plotting $A(T)$ for $T$, we can find the temperature where $A(T)=1$. The results are expressed in Figure \ref{AT}. The critical temperature obtained for $q=1$ is $T_{c}=0.69(2)$, which is slightly smaller than the literature value of $T_{c}=0.699$ and obtained by other methods. With the increase of the parameter $q$, as in the other methods, the critical temperature also decreases. The difference here is that we can't observe any change of behavior that indicates an extinction in the BKT transition or the emergence of a new class of phase transition.

%=====================================

%=====================================

\end{document}